\documentclass{article}

\usepackage{arxiv}

\usepackage[utf8]{inputenc} % allow utf-8 input
\usepackage[T1]{fontenc}    % use 8-bit T1 fonts
\usepackage{hyperref}       % hyperlinks
\usepackage{url}            % simple URL typesetting
\usepackage{booktabs}       % professional-quality tables
\usepackage{amsfonts}       % blackboard math symbols
\usepackage{nicefrac}       % compact symbols for 1/2, etc.
\usepackage{microtype}      % microtypography
\usepackage{lipsum}		% Can be removed after putting your text content
\usepackage{graphicx}
\usepackage{natbib}
\usepackage{doi}

\usepackage{tabularx}
\usepackage{booktabs}
\usepackage{array}
\usepackage{multirow}
\usepackage{enumitem}
\usepackage{xcolor}
\usepackage{tcolorbox} % https://www.tug.org/docs/latex/tcolorbox/tcolorbox.pdf#page=12.06
\tcbuselibrary{skins, breakable}
\tcbset{
  width=\textwidth,
  halign=justify,
  center,
  breakable,
  colbacktitle=gray,
  coltitle=white,
  fonttitle=\sffamily\small,%\bfseries
  colback=white,
  sharp corners, % arc=0mm,
  size=title,
  boxrule=0.1mm,
  fontupper=\small,%\ttfamily
}

\newlist{RQ}{enumerate}{1}
\setlist[RQ]{label=\textbf{RQ\arabic*.},ref={RQ\arabic*},itemsep=0ex,topsep=1pt}
\newcommand{\rqone}{How does a practitioner engage with a new domain in the era of GenAI?}
\newcommand{\rqtwo}{How does a practitioner integrate their prior knowledge with new domain knowledge in the era of GenAI?}
\title{Entering a New Domain in the Era of Generative AI: An Autoethnography of Solo Game Development as a Bioinformatics Educator}

\author{
    {Hannah Kim}\\
	Philadelphia, PA 19103 \\
	\texttt{https://hannahkimincompbio.github.io} \\
}

\date{September 10, 2026}

\renewcommand{\headeright}{}
\renewcommand{\undertitle}{}
\renewcommand{\shorttitle}{Entering a New Domain in the Era of Generative AI: An Autoethnography of Solo Game Development as a Bioinformatics Educator}

\hypersetup{
pdftitle={Entering a New Domain in the Era of Generative AI: An Autoethnography of Solo Game Development as a Bioinformatics Educator},
pdfsubject={human-computer interaction, bioinformatics},
pdfauthor={Hannah Kim},
pdfkeywords={generative AI, information search, interdisciplinary learning, informal learning},
}

\begin{document}
\maketitle

\begin{abstract}
	The rapid improvement of generative AI (GenAI) not only increased the efficiency of practice but also empowered individuals with capabilities to explore new domains. However, identity formation in the wake of such empowerment remains understudied. We use an autoethnography to understand identity formation in the context of solo game development as a bioinformatics educator. With the goal of developing a create-your-own-adventure (CYOA) game, we collected data for 73 days through journal entries, interacting with communities of practices. The autoethnographer had continuously reassessed the situation, and closely interacted with communities of practice. Their prior knowledge was used to provide in-game context and to uniquely interpret game development processes. These findings suggest that the autoethnographer made efforts to belong to communities of practice and engaged extensively in sensemaking through the lens of a bioinformatics educator. Although GenAI puts power at hand, attaining a new identity took place at a human pace.
\end{abstract}

% keywords can be removed

\section{Introduction}

As generative AI (GenAI) rapidly improves in performance, it is changing many facets of practice. Software engineers are shifting from manually writing code to reviewing AI-generated code to accelerate output delivery \cite{gropler2025future}. Designers co-create with GenAI across different phases of the design process \cite{cavallin2026designers}. Scientists can run discovery pipelines up to 200 times faster \cite{ghareeb2026multi} and maintain computational tools more effectively with GenAI \cite{Li2026AgenticAI}. Considerable effort is being devoted to exploring how GenAI can be further adopted.

In addition, GenAI has lowered barriers for non-experts to enter new domains. While an unbridgeable expertise gap still remains between domains separated by substantial knowledge distances, the very notion that crossing domains at will has become feasible with GenAI is monumental \cite{vendraminelli2025genai}. Indeed, there is not yet enough evidence in the literature to establish that GenAI capabilities can revolutionize interdisciplinary learning \cite{xia2026generative}. However, the pace at which GenAI models are improving has been so much faster than the publication cycle that these ideas may need to be retested.

In this work, we present an autoethnography of entering a new domain enabled by GenAI. For a practitioner of bioinformatics education, solo game development was unthinkable before GenAI, but advances in GenAI capabilities signaled the feasibility of entering the domain. Our sense of empowerment was justifiable in hindsight, especially since GenAI-assisted game development has been exploded in popularity around the time Anthropic's Fable 5 model was released during data collection \cite{Anthropic2026Fable5Mythos5}. However, we have yet to understand the implication of entering a new domain or its effects on newcomers' identities in the light of GenAI-induced sociotechnical changes.

The theoretical framework underlying our exploratory research questions is identity formation, with respect to both the new domain and the pre-existing practitioner identity. We investigate how participation in communities of practice influenced the autoethnographer's understanding of game development and how their prior knowledge influenced their sensemaking of information and unfamiliar challenges.

\begin{RQ}
\setlist[RQ]{leftmargin=0pt}
    \item \rqone
    \item \rqtwo
    % \item \rqthree
\end{RQ}

% [Contributions, overview of findings] refer to the CHI 2027 submission categories
Our contributions are (1) reflexive and (2) empirical. In presenting an autoethnography of a practitioner, we provide an analytical perspective on identity formation grounded in empirical observations of the autoethnographer and their interactions with communities of practice while collectively navigating a rapidly changing sociotechnical landscape. We capture a detailed personal account of entering a new domain and forming a new identity, a process that is becoming increasingly doable in the era of GenAI.

% \begin{enumerate}[leftmargin=15pt]
%     \item Opinion/Practical contribution?
%     \item empirical contribution
% \end{enumerate}

\section{Identity statement} \label{identity_statement}

The autoethnographer is a South Korean researcher in human-computer interaction based in the United States of America. They are bilingual and fluent in Korean and English, allowing them to access information from both Korean- and English-language sources. They are in their mid-30s, and are a practitioner of bioinformatics education who is interested in developing games as education tools. The particular genre of interest is create-your-own-adventure (CYOA) games. The autoethnographer likes playing video games, but has not developed a game outside of a computer science classroom as part of the curriculum. Throughout the manuscript, \textit{"we/us"} pronouns are used in accordance with academic convention. However, \textit{“they/them”} pronouns are used to refer to the autoethnographer (i.e., the author of this manuscript) during data collection, thereby maintaining analytical distance.

\section{Related works}

Educational games are digital games developed with the intention of providing entertainment and learning. Characterized as games with an additional goal of learning, educational games differ from gamification commonly found in education, which expands existing learning materials with game-based components and is therefore generally implemented on a smaller scale, \textit{"less than a full game"} \cite{intro2016serious}. While the high potential of using games for learning is well acknowledged by educators, the high cost of design and development, along with uncertainty about their effects on learner performance \cite{snyder2025enhancing, lameras2017essential, oliveira2026effects, de2011serious, schrader2023serious}, makes educational games challenging to adopt in the classroom \cite{intro2016serious}. The readily interdisciplinary nature of games demonstrates their potential for synergy with other disciplines, but the field of entertainment computing, a superset of serious games that is itself a superset of educational games, is fragmented precisely because different disciplines can drive the development of games \cite{intro2016serious}.

CYOA, chosen as the genre for our game development, traces its roots back to interactive children's gamebooks that were highly popular in 1980s America \cite{cook2021rearing}. From the very beginning, this format was considered more a game than a book \cite{cook2021rearing}, and it has been treated as a gamification strategy in education \cite{oliveira2026effects}. CYOA has consistently shown to improve learning experiences in science, technology, engineering, and mathematics disciplines \cite{tyndale2016keys, wettergreen2024meta, aguiniga2024effect, warning2021choose}. Despite its relative simplicity of implementation, CYOA activities are time-consuming to prepare \cite{wettergreen2024meta}. Clark C. Abt, who formalized the concept of serious games \cite{abt1987serious}, considered \textit{"cost-effectiveness"} of serious games relative to other instructional methods to be important for their adoption \cite{wilkinson2016brief}. From educators' perspectives, CYOA had not previously met that standard. With GenAI capabilities, serious games in the CYOA genre have the potential to become much more cost-effective and to develop into full-fledged games. However, this potential has not yet been actively explored in the literature.

Because identity is a lens through which the world is interpreted \cite{Brown2026ReflexivityPositionality,lencucha2025positionality}, identity formation is an important topic for practitioners. Identity formation is the continuous process through which individuals understand themselves so as to achieve internal coherence \cite{herman2011identity}. Although identity formation is typically used to discuss identity in terms of human personality, the concept has also been expanded to address both the process and the end product of identity in various contexts \cite{herman2011identity}, including professional identity formation (PIF). Identity formation is uniquely personal, requiring multiple iterations of sensemaking---managing \textit{"the cracks or gaps in individual cognition"}---in response to various situations newly encountered \cite{turner2023multifaceted}. However, it is also highly influenced by sociocultural aspects \cite{herman2011identity,tajfei1979integrative,waseem2008community,ABRAMS200114306}. The literature on PIF reflects these two aspects in detail: PIF inherently assumes pre-existing personal identities, which in turn further develop into new personal and professional identities through education and socialization, through which budding practitioners develop a sense of belonging to their communities \cite{reissner2024manifestations,sternszus2024being}.

Communities of practice provide opportunities for learning and socialization, even in informal settings. Person-to-person engagement with communities of practice fosters informal learning even without a structured curriculum, as members socialize \cite{johnson2022formal}. Informal learning is not organized around documented norms and standards; nevertheless, their amenability to individual learners' needs is advantageous \cite{johnson2022formal}. On the other hand, engaging with communities through social media may take different forms. Based on the categories of online communities suggested by Arthur Armstrong and John Hagel III, communities of practice fall under the communities of interest and relationship, allowing members to explore special topics and share experiences \cite{armstrong1998real}. Community members can in turn be categorized by stages of the membership lifecycle, with those in earlier stages described as peripheral and inbound based on participation \cite{wikipediaOnlineCommunity,lavewenger2008onlinecommunity}. An interesting way of learning through online communities is via lurking, a minimalist way in which members engage with content \cite{honeychurch2018learners,bozkurt2020lurking}. Comprising the vast majority, up to 90\%, lurkers only read and observe without evidence of participation \cite{nielsen2006participation}. These members may alternatively be interpreted as \textit{"apprentices"} of a community of practice; interestingly, despite their lack of visibility, lurkers were observed to have a sense of belonging \cite{honeychurch2018learners}.

Autoethnography is a qualitative research method that enables in-depth cultural understanding of the self and others through self-reflexivity \cite{chang2016autoethnography}. Distinct from ethnography, in which a researcher analyzes a culture as an external observer, autoethnography involves first-person analysis of a culture as an active member of a culture. Autoethnography is to be written carefully to maintain a balance between the individuality of the agent and the collective nature of culture, while recognizing that discovering the self within a culture is also a means of understanding others. Using autoethnography is beneficial in that data are already familiar to the researcher for analysis \cite{braun2023doing} and that the final report can be engaging to readers. It also allows the self and others to be better understood within a particular sociocultural moment. Finally, it can motivate the self and others to transform as a result of in-depth inquiry.

\section{Methods}

\subsection{Data collection and analysis}

As a beginner in game development, the autoethnographer immersed themselves in the domain by joining communities, experimenting with state-of-the-art GenAI systems, and designing a CYOA game intended to be both entertaining and educational. Data were collected through game development blog posts that functioned as a journal, documenting the development process, thoughts, and experiences. The study employed a first-person method (autoethnography), and no harm was experienced by the autoethnographer during data collection.

GenAI use was explicitly documented for every occurrence throughout the journal. To record anonymous but publicly accessible personal game development logs, Lovable was used to create the journaling website. During journaling, ChatGPT and Claude were used to assist with grammar and text editing. Figma and Computer Vision Annotation Tool (CVAT) were used to create non-AI images for the journal, while ChatGPT and Adobe Firefly were used in part to generate AI images. ChatGPT, Cursor, and Claude were used for game development planning and testing.

We used the six-step reflexive thematic analysis described by Braun \textit{et al.} \cite{braun2023doing} to analyze data. We systematically coded the entries, generated themes through an iterative process, and developed and refined these themes for the final report. Given that the data were collected as journal entries, reflexivity—the practice of critically examining one's experiences in relation to one's assumptions \cite{koopman2020autoethnography,Wikipedia2026ReflexivitySocialTheory,WarwickReflexivitySocialTheory}—was employed extensively during data analysis.

\subsection{Resource constraints}

This study was conducted under practical resource constraints. The autoethnographer was not a full-time game developer during the study period, resulting in limited time for game development. The autoethnographer also faced financial constraints and spent no more than \$20 per month on a GenAI system subscription. This expense represents a relatively modest consumer-level cost: in summer 2026, \$20 was approximately the cost of a decent lunch for one person in major cities in the United States.

\subsection{Sociotechnical context of GenAI}
The study is situated almost four years into the era of GenAI. Companies are competing globally for market share through continued GenAI model improvements, accompanied by heated discussions surrounding cybersecurity, regulation, and responsible use \cite{Anthropic2026ProjectGlasswing,Anthropic2026Fable5Mythos5,Anthropic2026RedeployingFable5,HuggingFace2026SecurityIncident,OpenAI2026HuggingFaceIncident,Anthropic2026CybersecurityIncidents,Microsoft2026OpenWeightModels,Amodei2026OpenWeights}. The competition among GenAI companies has also intensified demand for memory chips \cite{CNBC2026MemoryChipShortage}, contributing to supply constraints in 2026 that affect consumers, markets, and manufacturers, many of which are based in South Korea \cite{TechInsider2026MemoryChipShortage,Guardian2026AISelloffChipStocks}, the autoethnographer's country of origin.

\section{Results} \label{results}
% Positive or negative sentiment to GenAI?
% All sections should read like omnibus perspective of the same autoethnography. In each section, we provide analyses of the journal entries by themes.

\subsection{Collected data}
Over a period of 73 days---from May 24, 2026, to August 5, 2026---twenty-five journal entries totaling 14,844 words were collected, with an average of 593.76 words per entry (minimum: 291 words; maximum: 1,223 words), along with ten supporting images (See Figure \ref{fig:1}). These records provide a detailed account of a practitioner's entry into a new domain, offering a snapshot of the sociocultural phenomenon under investigation. The original journal entries are publicly accessible at: \url{https://cyoa-development-journal.lovable.app/read}.

\subsection{RQ1. \rqone}

\subsubsection{The autoethnographer continuously reassessed the context in which the game was being developed.} \label{beginner}

The autoethnographer shaped the design of their educational CYOA game by acquiring new information, thoroughly interrogating relevant knowledge, and updating priorities. They had to acquire knowledge about game development while simultaneously keeping pace with contemporary events. The summer of 2026 was marked by intense competition among AI companies for model performance, while the ethical, legal, and regulatory boundaries surrounding these technologies continued to change (See Figure \ref{fig:1}). Rapidly unfolding contemporary events further increased the need for frequent reassessment of context in the new domain.

\paragraph{Acquisition of new information}

The autoethnographer gained information about game development through direct and indirect experiences. They directly experienced planning game development workflows, writing stories, playing games, prototyping games, and processing images. Except for playing games, all of these activities involved GenAI to varying degrees. They learned about game design from the developer's perspective, such as why protagonists in many CYOA games \textit{“[sounded] so lost and forgetful in the beginning”} when establishing the initial context. By testing Anthropic's newly released Fable 5 model for prototyping, they learned that it could build a retro game with a zero-shot prompt in less than 10 minutes, making game development seem feasible even for a newcomer. In addition, they identified limitations in GenAI's storywriting capabilities; the narrative arc generated was still unsatisfactory when the system was provided only with the opening and closing. While direct experiences were occasional, indirect experiences were frequent. Indirect experiences were the autoethnographer's primary sources of information, and were formative to the game development. Instagram videos inspired the autoethnographer to make collages for in-game images, while reading Threads posts enabled them to identify a demand for educational games supporting memorization-intensive subjects. The communities of practice were major catalysts for indirect experiences (See Section \ref{communities_of_practice} for detail). Unlike their direct experiences, the autoethnographer supplemented none of these activities with GenAI.

\paragraph{Interrogation of relevant knowledge}

The autoethnographer habitually revisited and reflected on existing knowledge for game development in the context of the GenAI era. To provide an example, they repeatedly reflected on the meaning of creativity in the new era. Drawing on their everyday experiences of visiting museums, the autoethnographer compared the rise of Impressionism in the 19th century with the rise of GenAI and inferred that \textit{“heart-throbbing”} creativity was temporal. They realized that creativity in game development was also shaped by advancements of the time; retro games were small--only 31 kilobytes--indicative of the technological constraints of the 1980s, whereas contemporary games were substantially larger as technological advances enabled greater graphical complexity. They anticipated a similar pattern to occur with GenAI. Furthermore, they reflected on how creativity could be influenced by automation, whose repetitive nature is often considered antithetical to creativity. They questioned whether creativity could be interpreted as an ability to explore useful solutions within a space of possibilities and, hence, as inherently combinatorial and facilitable by automation. The autoethnographer observed that \textit{“this era [seemed to be making] a confession about how creativity \dots is in part incremental.”} They expressed excitement and concern about the implications of creativity achievable with GenAI. The concept of creativity itself was not new to the autoethnographer, yet they were strongly motivated to interrogate it anew as they entered the creative domain undergoing GenAI-induced changes.

\paragraph{Updating priorities}

The autoethnographer constantly sought to incorporate new information into the game development process, often resulting in updated priorities. Initially, the autoethnographer did not have the perspective of a game developer; they were clear about what they would do with the game's outcome, but they were mostly clueless about what was required to build one. As a first step, they drafted a game development workflow using GenAI and observed community examples to prioritize writing stories over supplying sounds and images. Upon immersing themselves in relevant experiences, the autoethnographer noticed that the typical reward system in CYOA games (i.e., collecting endings) could be difficult to implement in an educational game with a single learning objective; hence, designing the game around branching became their top priority. Anthropic's Fable 5 was released around this time. Until then, the autoethnographer had maintained a \$0 budget for computing resources, despite finding the situation—for example, relying on a daily free quota of five queries for vision models—unsustainable. With the release of Fable 5, however, they deprioritized this constraint because they experienced an extreme fear of missing out. The autoethnographer's game development relied on GenAI capabilities but was also potentially threatened by them had they set the bar too low. They ultimately relaxed the budget constraint because they needed to understand the GenAI experience available to their target players, attainable at a cost of \$20. Updating priorities thus reflected a deeper understanding of game development and its surrounding context.

\subsubsection{The autoethnographer's game development was closely informed by communities of practice.} \label{communities_of_practice}

As the autoethnographer noted, \textit{"[this] solo CYOA development, although done by one person, was likely not something that would be done in isolation." }Communities of practice were their primary sources of indirect experience. The autoethnographer engaged with game development communities both person-to-person and through social media, consistent with their goal of developing a game. Their practitioner identity also enabled person-to-person connections with other professional communities that overlapped in some aspects of practice (See Section \ref{identity_statement}): interaction designers, software developers, health sciences educators, and biomedical researchers. The local indie game developer community that the autoethnographer joined was identified through a social platform, whereas the other communities were extensions of the autoethnographer's existing social and professional connections. In addition, recommendation algorithms on their social media applications curated information from real-life connections and complete strangers who shared interests related to GenAI and the autoethnographer's identity. The comprehensive list of online social applications used during data collection was as follows: Bluesky, Facebook, Instagram, itch.io, LinkedIn, Reddit, Steam, Threads, Twine, X, and YouTube. Through engagement with these communities, the autoethnographer learned about community cultures and their reactions to GenAI.

\paragraph{Game development communities, person-to-person and through social media}

The autoethnographer closely observed game development communities to gain an understanding of cultural norms and community standards surrounding the use of GenAI. To acquire person-to-person experience and insights, they attended a playtest event hosted by a local indie game developer community, barely knowing more than a dictionary definition of the word \textit{"playtest"}. The event turned out to be an occasion for developers to pitch and demo their games and for community members to play new games and provide feedback. The autoethnographer enjoyed testing games and discussing the concept of fun with community members. They also learned about measures of success specific to game development (e.g., \textit{"being accepted to a gaming convention, being featured [in] a magazine, and releasing [a] game on Steam"}) as they listened to the developers' pitches. It was a valuable opportunity to learn about a game development community culture, but completely absent from the scene was any mention of GenAI, which was especially apparent to the autoethnographer, whose entry into game development had been incentivized by GenAI. When GenAI was so obviously permeating the domain, the lack of any signal about GenAI was itself informative: at the height of summer 2026, GenAI use was not something that a developer could openly acknowledge with pride in game development communities. In fact, the autoethnographer encountered many instances of fierce rejections of GenAI in game development communities as they wandered social media in search of information. Empathetically reflecting this sentiment, they brainstormed design alternatives that could align with it, such as using more conservative systems or Creative Commons Zero resources. The autoethnographer also became increasingly aware of different perspectives and practices regarding GenAI use in game development, but gauging the accepted boundaries of GenAI use within game development communities remained important to them throughout data collection.

\paragraph{Other communities of practice, person-to-person and through social media}

\begin{figure*}[ht]
    \centering
    \includegraphics[width=\textwidth]{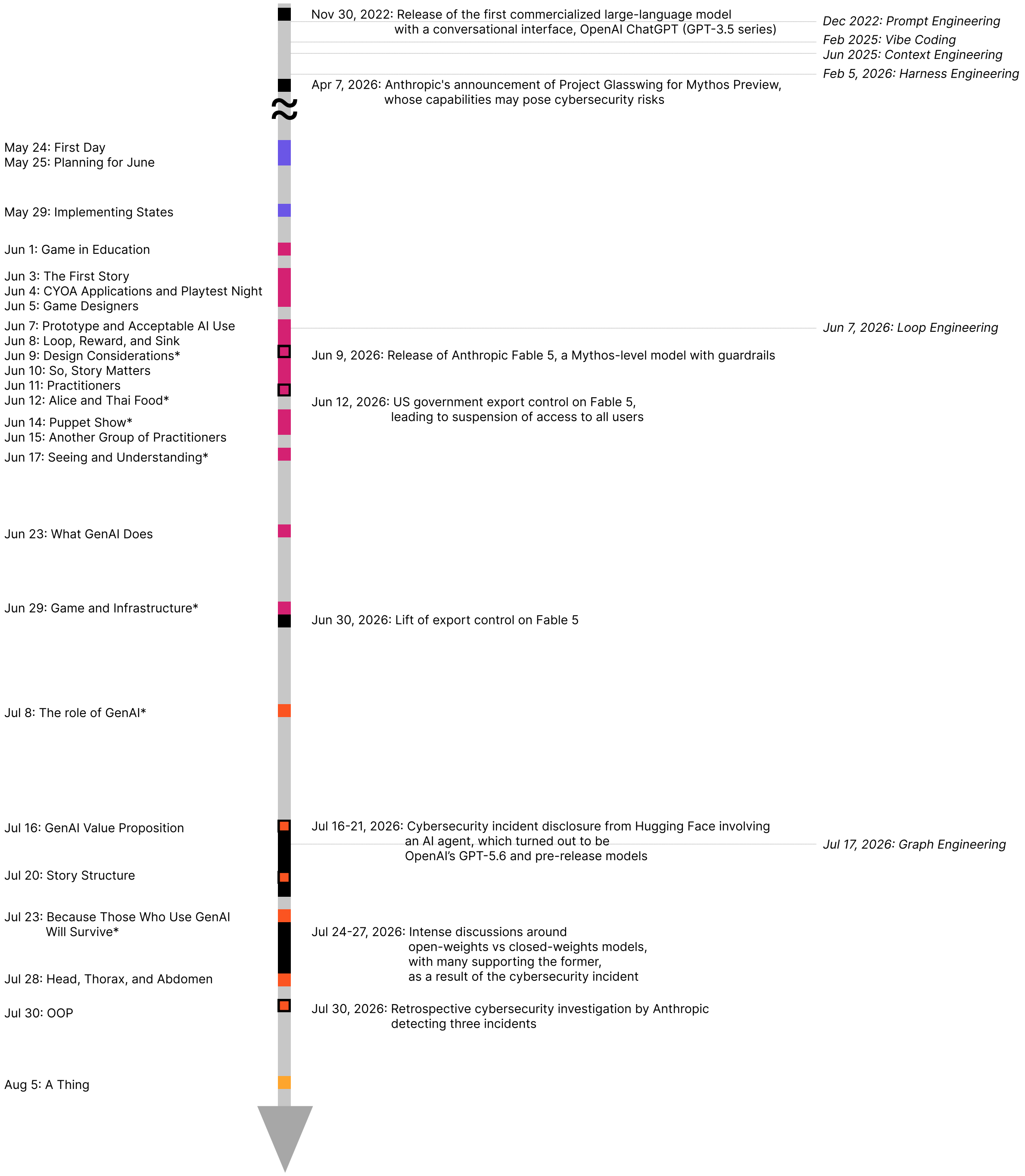}
    \caption{The Autoethnographer's Journal Timeline and Real-Time GenAI History Related to Game Development \cite{OpenAI2026ChatGPT,Anthropic2026ProjectGlasswing,Anthropic2026Fable5Mythos5,Anthropic2026RedeployingFable5,HuggingFace2026SecurityIncident,OpenAI2026HuggingFaceIncident,Anthropic2026CybersecurityIncidents,Microsoft2026OpenWeightModels,Amodei2026OpenWeights,Atlan2026HarnessEngineering,MITTechnologyReview2025ContextEngineering,PromptLayer2026ContextEngineering,Mitchell2026AIAdoptionJourney,Osmani2026LoopEngineering,Cherny2026ThreeStageLoop,LangChain2026GraphEngineering}. Asterisks (*) indicate entries with images.}
    \label{fig:1}
\end{figure*}

The autoethnographer gathered insights into developing a bioinformatics educational game befitting the era of GenAI from multiple communities of practice. These communities had collectively experienced the sociotechnical changes following the advancement of GenAI, but each had experienced them in distinct ways. The person-to-person community events that the autoethnographer attended had been organized for practitioners who leaned toward accepting GenAI. The interaction designers shared tips on how to use a \verb|SKILL.md| file to produce consistent outputs with reasonable token usage and how to precisely tweak the outputs to suit their design preferences by prompting for HTML outputs. The software developers assessed the cost of paying AI companies for API calls versus paying the electric bill for running a local GenAI system of similar performance at home, concluding the former to be more cost-effective. The health sciences educators focused their discussions on updating curricula for students, and they addressed how proper GenAI use could be taught during the didactic portion of training. Having been aware of the potential of AI for years before the rise of GenAI, the biomedical researchers were willing to adopt AI tools and were better prepared to ask for evidence of GenAI performance. To summarize, the practitioner communities carefully integrated GenAI-induced sociotechnical changes into their workflows to facilitate desired outcomes. Although the autoethnographer did not explicitly identify equivalent online communities of practice, they were not completely disconnected from these communities either. Social media recommendation algorithms often grouped them together based on overlapping interests in GenAI and professional practice, so the autoethnographer remained informed. Thanks to this connection, they were alerted to a series of cybersecurity-related events involving GenAI \cite{HuggingFace2026SecurityIncident, OpenAI2026HuggingFaceIncident, Anthropic2026CybersecurityIncidents, Microsoft2026OpenWeightModels, Amodei2026OpenWeights} that followed the U.S. government's export control of Fable 5 in response to security concerns \cite{Anthropic2026RedeployingFable5} (See Figure \ref{fig:1}). Community-level cybersecurity incidents involving commercial GenAI also ensued. The summer 2026 was an interesting historical period in which GenAI was capable of bypassing contemporary cybersecurity measures, while actions enabled by this capability carried mixed ethical and legal implications. They were also able to witness how new approaches to using GenAI systems were being introduced at increasingly shorter intervals \cite{Mitchell2026AIAdoptionJourney,Osmani2026LoopEngineering,Cherny2026ThreeStageLoop,LangChain2026GraphEngineering} (See Figure \ref{fig:1}). Despite these developments, the autoethnographer, as well as other practitioners, adhered to the harness engineering approach introduced earlier in the year, suggesting a potential asymmetry between the rapid development of GenAI technologies and their adoption in practice. Upon observing community responses and watching the 1995 film \textit{Ghost in the Shell}, the autoethnographer conceptualized the effect of GenAI on practitioner identities through the classic philosophical thought experiment of \textit{"the Ship of Theseus"}: if the activities around which a practitioner had built their identity are replaced by activities that produce comparable outcomes but involve vastly different processes, does the identity remain the same? Engaging with these communities provided not only an opportunity to learn about their perspectives but also to empathize with their sentiments and reflect on them.

\subsection{RQ2. \rqtwo} % Context of the CYOA game, of course, and also concepts frequently borrowed for sensemaking.

\subsubsection{The autoethnographer used their bioinformatics knowledge to provide context for storywriting and to interpret game development processes.} \label{bioinformatician}

The autoethnographer's identity as a bioinformatician helped them develop the CYOA game in both expected and unexpected ways. Because the autoethnographer had expertise in evolutionary biology and pandemic-related research, the game naturally developed a dystopian flavor. The goal of the game was for a player to successfully use a bioinformatics tool and learn how to interpret its output. To motivate an unsuspecting learner-player, the autoethnographer designed a protagonist who had survived twelve consecutive pandemics in a fictitious dystopia and was the only person eligible to conduct a bioinformatics analysis of the collected samples. In the story, the samples were automatically converted into data so that the focus could remain on bioinformatics. While entertaining the idea, they also bitterly titled the story \textit{"a suspicious future"}, given that bioinformatics analysis, despite its practical value for understanding how a pathogen evolved, was potentially more susceptible to automation than wet-lab analysis in reality. The autoethnographer was also able to immediately identify falsehoods in the analysis process when tentatively using underinformed GenAI to develop the story. While their knowledge was important in storywriting, it also became surprisingly useful as a lens through which they understood how a CYOA game should be designed and how GenAI should be perceived. In developing the game, the autoethnographer wondered how many small stories they would need to write, since each branching point would lead to at least two. Because the branching structure could be represented as a binary tree, the autoethnographer was able to estimate the maximum number of nodes (i.e., stories) for \textit{"a seemingly benign"} CYOA game with a depth of 10 using $2^{k+1}-1$, where $k=10$. Drawing on their prior knowledge of data structures, the autoethnographer realized the magnitude of labor required to implement—and eventually play—a CYOA game with a perfect binary tree structure and began to consider other, more feasible structures. In addition, their direct observation of tiger ocelli during a zoo visit, combined with a seemingly morphology-based argument from AI companies concerning the rise of intelligence in GenAI systems, prompted them to ponder the function and morphology of GenAI in the context of phylogenetic tree construction. In evolutionary biology, a tree can be constructed based on molecular building blocks or morphological characteristics of organisms, but not based on function, likely because function is context-dependent. AI companies' emphasis on artificial neurons in GenAI systems gave the impression that GenAI functions equivalently to biological neural systems, suggesting that similarity in structure had produced similarity in outcomes, including intelligence as operationally defined. Yet a closer look at this biomimicry alongside a typical phylogenetic tree of hominids reminded the autoethnographer that the comparison was based more on function than on morphology--much like how tiger ocelli can serve a function similar to that of actual tiger eyes in repelling predators. Because of this realization, they began to find the rampant use of this metaphor to promote GenAI performance distasteful, while still finding the performance impressive. Thus, just as the very goal of developing a CYOA game was, in hindsight, potentially influenced by the autoethnographer's familiarity with tree structures, their understanding of the new domain in the era of GenAI was also influenced by their background in bioinformatics.

\subsubsection{The autoethnographer's prior experience in education was important for assessing the utility of games in education, understanding the effects of GenAI on stakeholders, and maintaining the scope of the game design.} \label{educator}

The autoethnographer wanted to use games and GenAI to make bioinformatics more approachable to learners. Education is a vital process of gaining knowledge, but it is often practiced under constraints of time and budget. Hence, the autoethnographer also placed themselves under realistic resource constraints during game development. Because GenAI was largely capable even on free tiers, it seemed to be an accessible resource. The autoethnographer found GenAI to be a useful tool for game development, but also viewed the deterministic nature of branching points in CYOA games as a useful safety feature for managing learner-player inputs if GenAI were to be adopted more extensively in the game. Some game components were readily compatible with methods for quantifying performance used in education (e.g., multiple-choice questions and scoring). In addition, the element of repetition in games--or \textit{"looping"} in CYOA games--combined with rewards could work well for the repetition required in learning. To their disappointment, the autoethnographer found mixed evidence of improved performance through gamification in the literature but substantial evidence of improved engagement. To an educator, ensuring that learning had taken place was generally more valuable than promoting engagement. However, it was important to acknowledge what was feasible with games at an early stage; focusing on engagement (i.e., \textit{"fun"}) aligned well with the core values of game development communities. Even when engaging with communities, the autoethnographer's educator identity made them particularly perceptive to responses from learners and other educators regarding GenAI. About a year earlier, educators on social media had been excited about GenAI capabilities that enabled them to create visually appealing presentations using natural language. The autoethnographer no longer observed the same level of excitement, although it was clear that GenAI continued to be viewed positively by some. Among learners in fields traditionally known for memorization-intensive coursework, the autoethnographer detected neither clearly positive nor negative views on GenAI, but rather a decline in motivation as GenAI demonstrated strong capabilities for memorization and interaction with patients. Rote memorization had long been vilified by many, yet the autoethnographer came to a dramatic realization about its current inevitability when learning about context-free concepts not covered by common sense. Learning without motivation could only be challenging. Although the autoethnographer treated GenAI primarily as a convenient tool for game development, they realized that its existence could be highly demoralizing, especially for learners. Finally, the autoethnographer found their beliefs about learning shaping the design of the game's branching structure. For effective learning, the overarching goal had to be clear in the game. Such a goal would benefit from a skewed binary tree of branching points. Yet, the autoethnographer did not want to severely penalize the  potential for exploration when the game was already deterministic. How to turn less desirable choices into learning opportunities remained a work in progress, but the desire to maximize learning in players made it more likely that their CYOA game would, in part, take the shape of a balanced tree. The autoethnographer's understanding of the goals of education and perspective on its stakeholders crucially shaped the game's design.

\section{Discussion} % Title: Entering a New Domain in the Era of Generative AI: An Autoethnography of Solo Game Development as a Bioinformatics Educator

The results section revisits the collected journal entries four times, once for each analytical theme. Sections \ref{beginner} and \ref{communities_of_practice}, addressing RQ1, focus on the autoethnographer's experience as a new member entering the domain of game development amid rapid sociotechnical changes introduced by GenAI, while Sections \ref{bioinformatician} and \ref{educator}, addressing RQ2, focus on their experience as a bioinformatics educator entering the domain. The results indicate that the autoethnographer's formation of a new identity as a game developer involved both external and internal processes.

\subsection{RQ1. \rqone}

RQ1 situates the autoethnographer's identity relative to the domain. The autoethnographer is new to the domain as much as the domain is new to the autoethnographer, and their prior domain knowledge is miniscule. Even though the autoethnographer equips themselves with the capabilities of the ever-friendly GenAI, their primary source of information is communities of practice, both person-to-person and through social media. Information obtained in this way is often raw, open to interpretation, and without accumulated legitimacy, but it is fast and current, particularly within the rapidly changing technological landscape surrounding GenAI models. As a newcomer, the autoethnographer's assessment of information takes three steps: acquiring new information, interrogating relevant knowledge, and updating priorities. The need for constant reassessment is a hallmark of managing information whose value has not been well established. The autoethnographer is metacognizant of their insufficient domain knowledge and hence cautious.

Navigating the new domain, the autoethnographer used GenAI selectively. They used GenAI to plan game development workflows, write stories, prototype games, and process images. These were direct components of game development that informed the autoethnographer about details of why certain design decisions made more sense and what could go wrong. Barriers to these activities were significantly lowered by GenAI, thanks to its capabilities and the sense of empowerment it provided. On the other hand, the autoethnographer did not use GenAI to play games, engage with posts on social media, review literature, interrogate the meaning of creativity or decide what to do next. Many of these activities could have been automated with GenAI, had the autoethnographer been willing to do so. Yet, the autoethnographer did not feel the need to use GenAI for these activities because they readily enjoyed them or felt subjectively competent at them. Perhaps this is an important distinction. The autoethnographer's use of GenAI depended on their own assessment of whether they could or wanted to perform an activity, rather than on whether GenAI could perform it. This observation suggests a rationale for learning in an era where GenAI is becoming increasingly capable and ubiquitous: even if GenAI lowers barriers to entering domains in previously unimaginable ways, one still needs to learn about oneself.

The autoethnographer's new identity draws inspiration from communities of practice. They consider a domain to be inherently inseparable from its communities. Community culture provides insight into why and how domain knowledge develops. Cultural norms and community standards constitute a trove of heuristics, representing the accumulated memory of experiences. As a result, to benefit from existing knowledge and build upon it, the autoethnographer found it natural to belong to communities. The era of GenAI that enabled the autoethnographer to enter game development was also one of the most tumultuous periods for the identity of practitioners. The sociotechnical changes implied potential changes to practices, which were themselves identity-defining. Consequently, game development communities were in the process of establishing a new standard for what constituted legitimate practice, with negative sentiments more strongly represented at the time. Through the lens of \textit{"the Ship of Theseus"}, an identity may depend on the continuity of the original concept or the continuity of its components, and game development communities were seeking a balance between the two. Other communities of practice that the autoethnographer engaged with had also been influenced by the sociotechnical impacts of GenAI, and these communities were generally more inclined to accept GenAI. An asymmetry existed across different domains between the pace of societal adoption and that of technical development, but GenAI was undeniably practice-changing.

Belonging to communities of practice is important for the formation of the autoethnographer's new identity. To belong is to identify cultural boundaries of these communities and to adhere to them. Not only does the autoethnographer follow the design conventions of game development communities, but they also consider design options that reflect awareness of negative sentiment toward GenAI image generation--for example, leaning toward more conservative systems and Creative Commons Zero resources. Games are an intuitive and widely approachable medium for testing the improved capabilities of new GenAI models, and the availability of GenAI has consequently made entry into game development much easier. However, that is not to say that everyone who enters would be capable of developing a sustained affinity for game development sufficient to succeed or improve upon domain knowledge over time, as members of these communities have done. Being enabled to enter the domain was not itself identity-altering, and that is why the autoethnographer wanted to belong: to form an identity.

\subsection{RQ2. \rqtwo}

RQ2 examines the autoethnographer's newly forming identity in relation to their existing identities. Identity formation is sociocultural, but it is also deeply personal. The autoethnographer was new to the domain, but they were not without pre-existing identities. They were a new game developer, but they were also a bioinformatician and educator. Although these identities seemingly had little overlap with a game developer identity, they influenced the design of the game and eventually the perspective through which game development made sense to the autoethnographer.

The autoethnographer's pre-existing identities serve as a starting point for game development. With a narrative-intensive game in mind, the autoethnographer's in-depth understanding of bioinformatics was crucial for designing the fictional world in which the game takes place. Each story had to be educational, reasonably credible, and intriguing. Particularly when using GenAI, the quality of both storywriting and game prototyping depended on the autoethnographer's identity as a bioinformatician, because incongruities could not otherwise be identified. On the other hand, as an educator, the autoethnographer identified educational values in developing a game. For example, GenAI could be used in the game, but potential harms had to be mitigated. Resource use required for gameplay had to be practical. Incentives for repetition in games had be leveraged for learning. And so forth. These values guided how the game was to be developed and what values could not be compromised from an early stage. The autoethnographer's prior knowledge was useful in their game development because the goal itself was strongly aligned with their identities as a bioinformatician and educator.

The autoethnographer uses their prior knowledge to make sense of the conceptual gaps that they identified while developing games with GenAI support. The autoethnographer had a rough idea of what an enjoyable CYOA tree depth was and how time-consuming storywriting could be. However, they lacked intuition about what this implied for their own game development. Therefore, the autoethnographer compensated for this gap as a bioinformatician would: by simplifying the problem and applying data structure approaches. Because they viewed the game as a bioinformatician, the autoethnographer has made better sense of how to structure the game. In addition, the autoethnographer developed a better understanding of GenAI as a tool through the lens of bioinformatics. While pondering the anthropomorphized roles of GenAI amid the overflowing news on game development, the autoethnographer attempted to place GenAI in the phylogenetic tree of hominids, only to fail. This exercise disillusioned the autoethnographer with the lifelike wonder that GenAI was packaged with by the virtue of biomimicry of neural networks, enabling them to make more concrete sense of GenAI as a useful game development tool. As an educator, the autoethnographer was able to make sense of how commercial educational games cleverly provided context for otherwise context-free memorization and realized the real danger that low motivation poses to learning in the presence of GenAI. Overall, the autoethnographer was internally developing a new identity alongside a game, and pre-existing identities supplemented the formation of it.

\section{Conclusion}

This autoethnography analyzes the process of entering a new domain while already being a practitioner of another domain. Data collected over the summer of GenAI-induced sociotechnical changes vividly record the autoethnographer's scramble to develop a game, both as a member of different communities of practice and as a bioinformatician and educator. GenAI enabled the autoethnographer's entry into game development, but the process they went through indicates that attaining a new identity is much more complex and takes place at a human pace. In the era of GenAI, we expect to see more and more people make ambitious attempts to enter different domains, discover the previously unimaginable, and build new, unique identities. We hope this autoethnography provides a detailed example of what it was like to undertake such an attempt during this exciting and chaotic period.

\section{Limitations}
Community gatherings are rarer during summer. More frequent person-to-person engagement with communities may have been achieved with a longer period of data collection. However, the autoethnography provides a interpretive snapshot of a tumultuous period in which GenAI underwent rapid advancement and broadly impacted multiple communities. Hence, despite the limited opportunities for person-to-person engagement, the data provide a meaningful account of the autoethnographer's experience during this period.

%% Figures
% Figure description should be less than 2000 characters including spaces.

%%
%% The acknowledgments section is defined using the "acks" environment (and NOT an unnumbered section). This ensures the proper identification of the section in the article metadata, and the consistent spelling of the heading.
\section*{Acknowledgments}
We would like to acknowledge the practitioner communities that the autoethnographers were a part of. It was because of the open discussions in these communities that we were able to write this paper. The communities of software developers, health sciences educators, and biomedical researchers are based in Philadelphia, PA, and the community of interaction designers is based in San Francisco, CA.

\bibliographystyle{unsrt} % abbrvnat or unsrt
\bibliography{chi2027-x}  %%% Uncomment this line and comment out the ``thebibliography'' section below to use the external .bib file (using bibtex) .

\end{document}